\documentclass[12pt]{iopart}
\usepackage[]{graphicx}
\begin{document}
\title[Floquet/Bloch-band fusion phenomenon]
{Note on a Floquet/Bloch-band fusion phenomenon in scattering by truncated periodic multi-well potentials}
\author{K-E Thylwe }
 \address{Kl\"{o}verv\"{a}gen 16, 387 36 Borgholm, Sweden \\
(Retired from Department of Mechanics, KTH- Royal Institute of Technology)\  }
\ead{kethylwe@gmail.com}

\begin{abstract}
A transmission phenomenon for a  quantal particle scattered through a multi-well potential in one dimension is observed by means of an amplitude-phase method.  The potential model consists of $n$ identical potential cells, each containing a symmetric well. Typical transmission bands contain $n-1$ possible energies of total transmission. It is found that certain band types contain $n$ energies of total transmission. A fusion phenomenon of this type of band with a typical neigboring band is also found. As the transmission gap between them collapse and disappear, a resulting fused single band is seen to contain $2n-1$ energy peaks of total transmission.
\end{abstract}


\section{Introduction and presentation of results}
Basic quantum physics related to truncated periodic potentials focuses on transmission and reflection properties. A free particle wave enters the potential region from one side (here from the right side); part of the wave is reflected (here in the direction $x\to+\infty$), while the remaining part of the wave is transmitted through the potential region (here in the direction $x\to-\infty$). From intensities of wave component one obtains the so-called reflection and transmission coefficients.

The present approach is similar to that for wave scattering by a single potential barrier or well \cite{T05a}. A main difference in this study is that the interaction has a finite periodic structure of potential 'cells', where each cell contains a symmetric well. This part of the potential is treated as described in \cite{T19b}.

Quantal transmissions through multi-barrier/well systems in one dimension \cite{Griffiths01}-\cite{Bar05} provide important theoretical notions for analyzing tunneling in solids \cite{Duke69}, chemical selection of gas components \cite{Mandrˆ14}, electronic properties of material structures such as graphene \cite{ReviewsG}-\cite{Kumar18}.  Generalizations involve $2D$ models, relativistic corrections, and spin-coupling effects \cite{Pyykko88}-\cite{Chen16}. Transmission effects caused by various non-vanishing exterior potentials are also in progress \cite{T20a}.

The truncated periodic potential is assumed vanishing outside an interval $0\leq x\leq n\pi$, where $\pi$ is the dimensionless unit length of the period (or cell) and $n$ is the number of such cells. The potential satisfies $V(\pi)=V(0)=0$ and its derivative with respect to $x$ satisfies $V'(\pi)=V'(0)=0$. The analytic form of the multi-well potential used  for numerical illustrations is
\begin{equation}
V(x)= V_0 \sin^4(x),\;\; 0\leq x\leq n\pi,\;\;V(x)=0,\;x<0, x>n\pi, \label{poti}
\end{equation}
where $V_0<0$ is the well depth. 
Figure 1 shows a potential with 6 identical wells in the range $0\leq x \leq 6\pi$.  Arrows indicate directions of propagating quantal waves. 
\begin{figure}
\begin{center}
\includegraphics[width=80mm,clip]{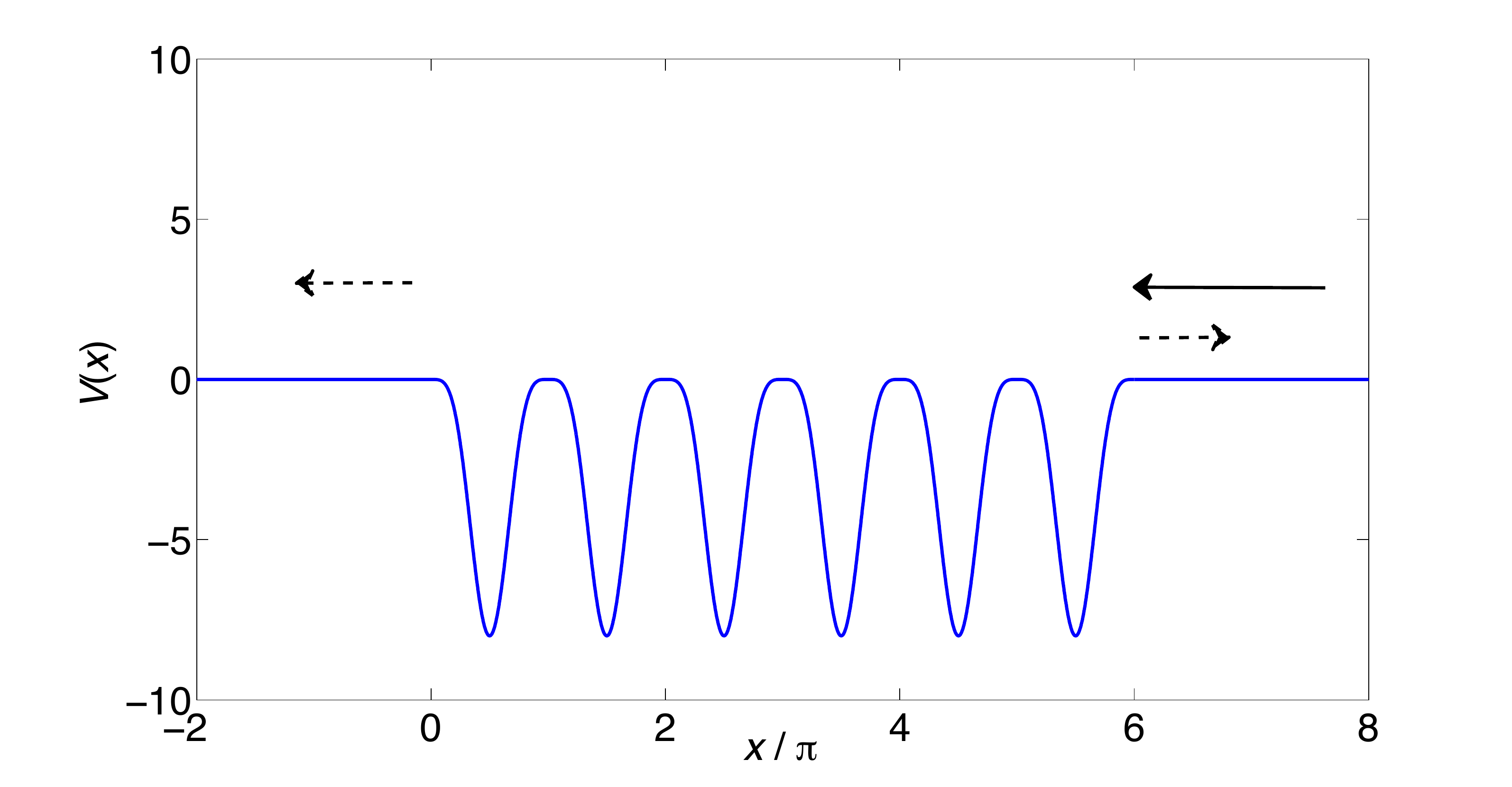}
\end{center}
\caption[ ]{\label{fig1} \small Illustration of a multi-well potential and traveling wave components. The incoming wave component (thick arrow) enters from the right.}
\end{figure}

The time-independent Schr\"{o}dinger equation with dimensionless parameters is expressed as
\begin{equation}
\frac{\rmd^2 F(x)}{\rmd x^2} + 2\left[E-V(x) \right] F(x) =0,
\label{PO} 
\end{equation}
as if presented in atomic units.
The symbol $E (>0)$ represents the total scattering energy.  Equation (\ref{PO}) is a special case of a so called Hill equation \cite{Hill}-\cite{hay:oscil} for $V(x)$ in (\ref{poti}).
Scattering boundary conditions for the wave function $F(x)$ are
\numparts\begin{eqnarray}
F(x) \sim t \frac{1}{\sqrt{k}}\exp (-\rmi kx), && \;\; x \rightarrow - \infty,\,\,\, \label{leftB}\\
F(x) \sim \frac{1}{\sqrt{k}}  \exp (-\rmi kx)&&
+ \,\,\, r \frac{1}{\sqrt{k}} \exp (\rmi kx), \;\; x \rightarrow + \infty,
\label{rightB}
\end{eqnarray}
\endnumparts
where $t$ and $r$ are the transmission and reflection amplitudes, respectively. The reduced (angular) wave number is
\begin{equation}
k=\sqrt{2E}.
\end{equation}
The complex-valued amplitudes  $t$ and $r$ determine  the transmission and reflection coefficients from
\begin{equation}
T=|t|^2,\;\;R=|r|^2.  \label{RTdef}
\end{equation}

The main results are presented next. Derivations by using an amplitude-phase method introduced recently in \cite{T19b} are deferred to a subsequent section. Original ideas of representing a quantal wave in terms of an amplitude function and a phase function are presented in \cite{Pinney}-\cite{Young}. 

The energy dependence of the transmission coefficient within a (Floquet/Bloch) band is analyzed in some detail from an exact expression
\begin{equation}
T(E) = \left(1+ \left| \Lambda(E)\right|^2 \right)^{-1},
\end{equation}
containing an imaginary quantity 
\begin{equation}
\Lambda(E) = \rmi J(E) \sin n\alpha(E).  \label{LambdaLP}
\end{equation}
$\alpha(E)$ is the real Floquet/Bloch phase as function of energy $E$, an intrinsic phase defined by the wave propagation across a single cell. This phase is independent of which exact method is used. The energy dependent factor $J(E)$ is independent of $n$ and is singular at band edges. Total transmission occurs at zeros of $\Lambda(E)$, i.e. by either of the conditions
\begin{eqnarray}
J(E_J) = 0, \\  \alpha_{nj\nu} = (j+\nu/n)\pi,\;\;\alpha_{nj\nu}=\alpha(E_{nj\nu}). \label{FBcond}
\end{eqnarray}
A band number is represented by $j=0, 1, \cdots$, and $\nu = (0), 1, \cdots, n-1$ counts the energies of total transmission within a given band.
Total transmission caused by a zero of $J(E)$, for $E=E_J$, is assigned an index '$J$' instead of a $\nu$-number.  $J(E)$ is not related to the much 'faster' Floquet/Bloch phase $\alpha$, as function of energy. Zeros of $J(E)$ are not present in all bands. Single zeros occur in particular bands.

The case $\nu=0$ is usually forbidden in a typical $j$-band.  There, the phase $\alpha(E)$ is confined to an interval $j\pi\leq \alpha(E) \leq (j+1)\pi$. Phase values $\alpha=j\pi$ and $\alpha=(j+1)\pi$ correspond to band edges, where $J(E)$ is singular and $\sin n\alpha(E)$ is zero. Note that the product of these quantities, and $\Lambda(E)$, is still finite.  
If two neighboring bands, e.g. the $j$- and $(j+1)$-bands, fuse, the case $\nu=0$ becomes valid for the $(j+1)$-band. Then, $\alpha$ is real in the larger interval $j\pi\leq \alpha(E) \leq (j+2)\pi$. This fusion phenomenon is illustrated in figures 1 and 2.

\begin{figure}
\begin{center}
\includegraphics[width=100mm,clip]{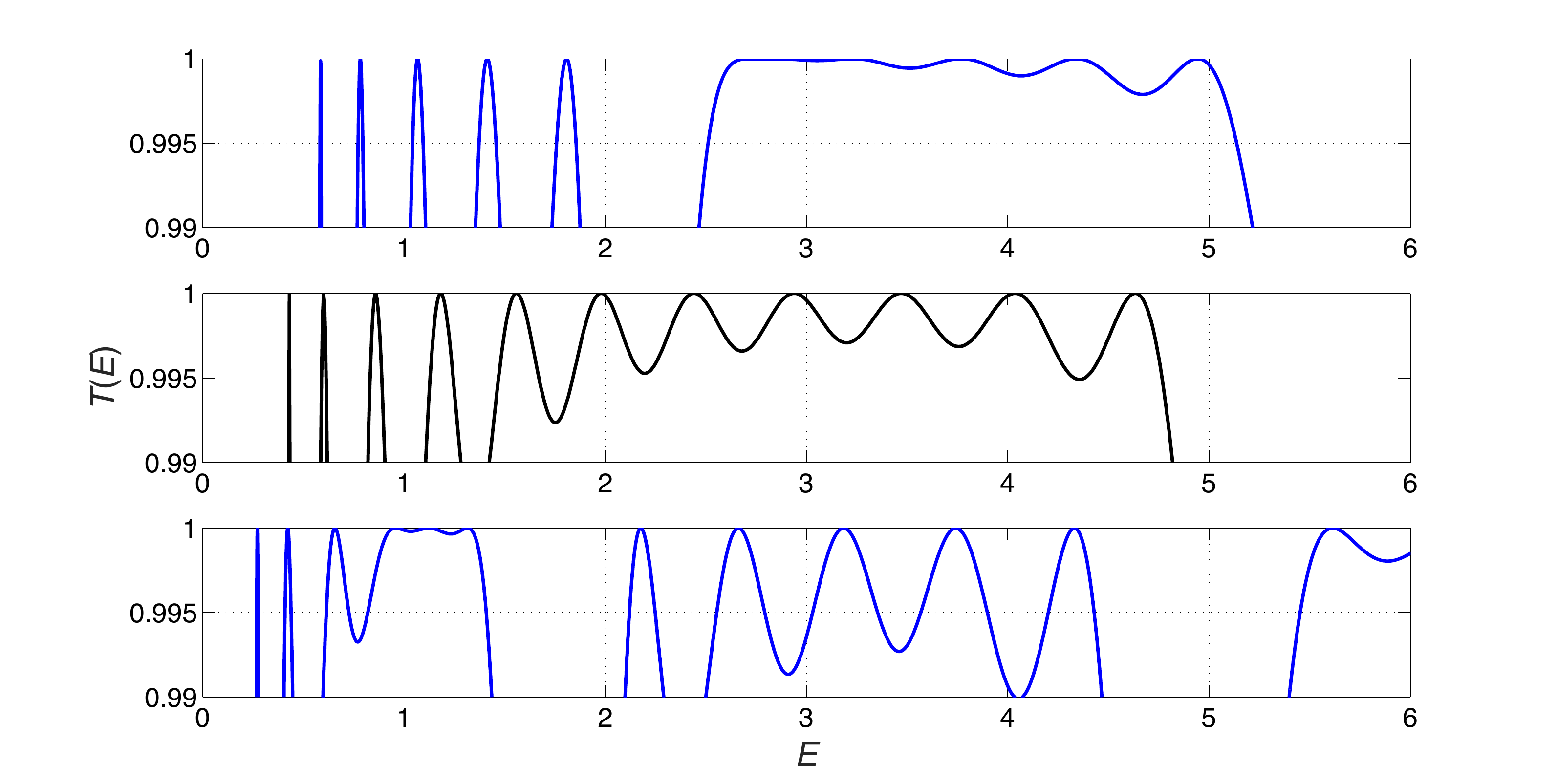}
\end{center}
\caption[ ]{\label{fig1} \small  Energy behaviors of  $T$ for potential parameter values $V_0=-7$ (top subplot), -8 (middle subplot), and -9 (bottom subplot).  Two transmission bands with $j=2$ and 3 are separated  in the top and bottom subplots. In the middle subplot, the fused band with $j=(2,3)$, there are 11 peaks of total transmission. Only one of two of the separated transmission bands shows 5 simple peaks of total transmission. The neighbor band has a broad peak structure which obscures the sharper peaks.}
\end{figure}
\begin{figure}
\begin{center}
\includegraphics[width=100mm,clip]{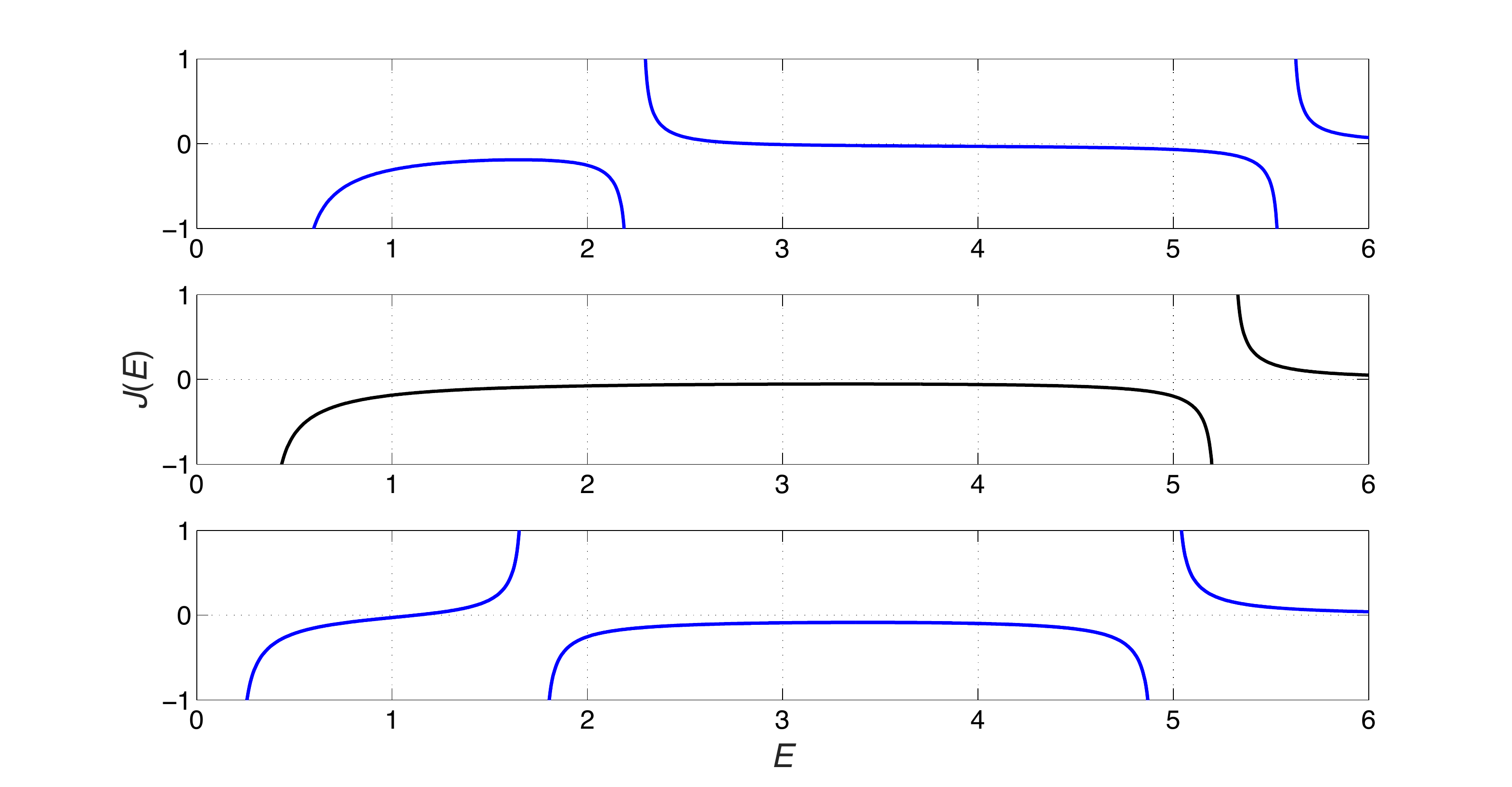}
\end{center}
\caption[ ]{\label{fig1} \small Energy behaviors of $J(E)$ for potential parameter values $V_0=-7$ (top subplot), -8 (middle subplot), and -9 (bottom subplot).  $J(E)$ has a zero passage near $E=2.85$ for $j=3$  in the top subplot, and near $E=1.12$ for $j=2$ in the bottom subplot. In the middle subplot, the fused band $j=(2,3)$, $J< 0$.}
\end{figure}
Figure 1 shows transmission coefficients as functions of energy for $n=6$ and three potential parameter values $V_0=-7, -8$, and -9. The two first bands, $j=0, 1$, have negative energies.
The top subplot ($V_0=-7$) shows two separated transmission bands corresponding to $j=2$ and $j=3$. The number of energy peaks of total transmission is 5 in the band $j=2$. Two peaks in the band $j=3$ are less sharp. A detailed analysis shows that the peak at $E\approx2.85$ is due to the single zero of $J(E)$. A close-lying peak at $E\approx2.72$ is due to the phase factor $\sin n\alpha(E)$ in equation (\ref{LambdaLP}). There are 6 peaks of total transmission in band $j=3$.

The bottom subplot ($V_0=-9$) in Figure 1 shows that the band with $j=3$ has 5 peaks of total transmission. For $j=2$, two peaks near the band edge are unclear. A detailed analysis shows that the peak at $E\approx1.12$ is due to the single zero of $J(E)$. A close-lying peak at $E\approx1.32$ is due to the factor $\sin n\alpha(E)$ in equation (\ref{LambdaLP}).
Other peaks of total transmission are due to the phase factor $\sin n\alpha(E)$. 

In the middle subplot ($V_0=-8$) all peaks of total transmission are due to zeros of $\sin n\alpha(E)$. $J(E)$ has no zeros. The total number of peaks is $2n-1$. The way a gap between such bands vanishes is explained in detail in \cite{T20a}.

Figure 2 illustrates the energy behaviors of $J(E)$ in the complete band zones $j=2$ and $j=3$. $J(E)$ is imaginary in gap zones and is not illustrated. The top subplot shows that $J(E)$ has a zero and changes sign within band $j=3$. The bottom subplot shows the same thing except that the zero occurs in band $j=2$. The middle subplot shows the fused band, where $J(E)<0$.

Energy peaks of total transmission seen in the middle subplot of Figure 1 are computed and displayed in Table 1.  The fused transmission band contains $2n-1 = 11$ such energies. Both quantum numbers, $j=2$ an 3, are used. The quantum number $\nu=0$ appears for $j=3$. All energies are due to the Floquet/Bloch phase condition in (\ref{FBcond}). 
\begin{table}
\begin{center}
\small
\begin{tabular}[t]{rrcc} \hline
\hline
\hline
&$(n,j,\nu)$&&$E_{n,j,\nu}$($V_0=-8$) \\
\hline
\hline
\hline
\hline
&$(6,2,1)$&&0.4310069\\
&$(6,2,2)$&&0.6020104\\
&$(6,2,3)$&&0.8594292\\
&$(6,2,4)$&&1.1829404\\
&$(6,2,5)$&&1.5592963\\
\hline
&$(6,3,0)$&&1.9804654\\
\hline
&$(6,3,1)$&&2.4414865\\
&$(6,3,2)$&&2.9392161\\
&$(6,3,3)$&&3.4715098\\
&$(6,3,4)$&&4.0368108\\
&$(6,3,5)$&&4.6336285\\
\hline
\hline
\end{tabular}
\caption{\small Energies $E_{n,j,\nu}$ of total transmission for $V_0=-8$.}
\end{center}
\label{table1}
\end{table}
\begin{table}
\begin{center}
\small
\begin{tabular}[t]{rrcc} \hline
\hline
\hline
$(n,j,\nu)$&&$E_{n,j,\nu}$($V_0=-7$)& $E_{n,j,\nu}$($V_0=-9$)\\
\hline
\hline
\hline
$(6,2,1)$&&0.5864576&0.2720599\\
$(6,2,2)$&&0.7839609&0.4232600\\
$(6,2,3)$&&1.0683904&0.6575990\\
$(6,2,4)$&&1.4144690&0.9593766\\
$(6,2,J)$&& $--$ &1.1245588\\
$(6,2,5)$&&1.8083837&1.3158913\\
\hline
$(6,3,1)$&&2.7205677&2.1769876\\
$(6,3,J)$&&2.8508686&$--$\\
$(6,3,2)$&&3.2275311&2.6622808\\
$(6,3,3)$&&3.7680369&3.1848282\\
$(6,3,4)$&&4.3405175&3.7418530\\
$(6,3,5)$&&4.9438598&4.3311594\\
\hline
\hline
\end{tabular}
\caption{\small Energies $E_{n,j,\nu}$ and $E_{n,j,J}$ of total transmission for $V_0=-7$ and $V_0=-9$. The symbol '$J$' is used for peak energies caused by $J(E)$.}
\end{center}
\label{table2}
\end{table}

Table 2 contains energy peaks of total transmission seen in  Figure 1 for the top and bottom subplots. Here, the transmission bands $j=2$ and 3 are separated by a gap zone. Peaks of total transmission due to $J(E)$ occur. The total number of peaks for $j=2$ and 3 is 11, the same as in Table 1, but the reasons are different.

If exterior potentials are added to the trucated potential considered , the fusion phenomenon remains. Results of a more general investigation will be published esewere \cite{T20a}.
\section{Derivations}
To obtain $t$ and $r$ in (\ref{leftB}) and (\ref{rightB}), particular amplitude-phase solutions in each characteristic region are introduced. There are two asymptotic regions and the region of $n$ identical cells. 

Two independent solutions of (\ref{PO}) are defined in terms of a positive amplitude function $A(x)$ and a related real phase function $p(x)$ as \cite{T19b} 
\begin{eqnarray}
\Psi^{(\pm)}(x)=A(x) \exp (\pm\, \rmi \, p(x)),\label{ansatz}\\ p'(x) = A^{-2}(x) \;(>0), \label{Mephase}
\end{eqnarray}
where $'=\rmd/\rmd x$.
Due to the relation (\ref{Mephase}), the Wronskian determinant of the two solutions (\ref{ansatz}) is independent of $x$ \cite{T05a}. 
Any amplitude function satisfies a nonlinear Milne-Pinney equation \cite{Milne}-\cite{Pinney}
\begin{equation}
\frac{\rmd^2 A(x)}{\rmd x^2} +2\left[E-V(x) \right]  A(x)={A}^{-3}(x).
\label{Me} 
\end{equation}
Amplitude functions differ by their boundary conditions \cite{APbound,T18b}. For any choice of $A(x)$ one has two independent exact solutions $\Psi^{(\pm)}(x)$.  An amplitude function is known to be more or less oscillatory due to different choices of its boundary conditions. Several amplitude functions may be used to represent a given linear wave function. Different representations of a linear wave function can be expressed in terms of the others by linear combinations.

Equation (\ref{Me}) is re-written for computational purposes as a first-order differential equation as
\begin{equation}
\left[\begin{array}{c}   A(x)  \\     A'(x) \\   p(x)  \end{array} \right]' 
= \left[ \begin{array}{c}    A'(x)  \\  {A}^{-3}(x)- 2(E-V(x)) A(x) \\  A^{-2}(x) \end{array} \right]. \label{num}
\end{equation}
The integration starts from boundary conditions of the amplitude function. The phase function needs a specified integration constant.

Amplitude-phase solutions $A(x)$ of (\ref{Me}) are used locally, in each characteristic region of $x$ \cite{APbound,T18b}. Firstly the two exterior regions are considered. 
The two exterior solutions, with amplitude functions $A_{L}(x)=A_{R}(x)=k^{-1/2}$ of (\ref{Me}), are:
\numparts\begin{eqnarray}
\Psi_{L}^{(\pm)}(x)=k^{-1/2} \exp (\pm\, \rmi kx).
\label{phiL}
\\
\Psi_{R}^{(\pm)}(x)=k^{-1/2} \exp (\pm\, \rmi kx)  \exp (\mp\, \rmi \,kn\pi), \label{phiR}
\end{eqnarray}\endnumparts
$x=0$ and $x=n\pi$, are reference points for the respective phases. 
Fundamental solution matrices consist of $\Psi_{L,R}^{(\pm)}(x)$ in the upper row and $\Psi'^{(\pm)}_{L,R}(x)$ in the lower row. 
The exterior fundamental solutions satisfy
\begin{equation}
{\bf \Psi}_L(0) =\left(\begin{array}{cc} k^{-1/2} & k^{-1/2}\\    \rmi k^{1/2} & - \rmi k^{1/2} \end{array}\right),\;{\bf \Psi}_R(n\pi) =\left(\begin{array}{cc} k^{-1/2} & k^{-1/2}\\    \rmi k^{1/2} & - \rmi k^{1/2} \end{array}\right).\label{LRPsi0}
\end{equation}

In the region of the locally periodic potential it is convenient to use a real-valued fundamental solution matrix composed by the real and imaginary parts of (\ref{Mephase}). A further simplification results from the use of a periodic amplitude function $A_p(x)$. Such an amplitude function is defined by particular boundary conditions at the first cell boundary point $x=0$, chosen as \cite{T19b}
\begin{equation}
A_p(0) = u_p, \;\; A'_p(0) =0. \label{bcondM}
\end{equation}
The corresponding phase satisfies $p'_p(x)=A_p^{-2}(x)$, and the phase reference point is taken at $x=0$. Particular phase values are 
\begin{equation}
p_p(0) = 0,\;\; p_p(\pi)=\alpha;\;\;  p_p(n\pi)=n\alpha.
\end{equation}
A principal fundamental solution matrix is defined by the solutions
\begin{equation}
S(x)=A_p(x)/u_p\sin p_p(x),\;\;C(x)=A_p(x)u_p\cos p_p(x). \label{apSCsol}
\end{equation}
 as
\begin{equation}
{\bf \Psi}(x) = \left(\begin{array}{cc}
   C(x) & S(x) \\ 
   C'(x)& S'(x)\\ 
 \end{array}\right),\;\; \det {\bf \Psi}(x) =  1,  \label{PsiM}
\end{equation}
satisfying
\begin{equation}\fl
{\bf \Psi}(0) = \left(\begin{array}{cc}
   1 & 0 \\ 
   0 & 1
  \end{array}\right),\;{\bf \Psi}(\pi) = \left(\begin{array}{cc}
   \cos \alpha &u_p^2 \sin \alpha \\ 
   -u_p^{-2} \sin \alpha &\cos \alpha
  \end{array}\right),\;{\bf \Psi}(n\pi) = \left(\begin{array}{cc}
   \cos n\alpha &u_p^2 \sin n\alpha \\ 
   -u_p^{-2}\sin n\alpha & \cos n\alpha
  \end{array}\right).   \label{PsiMn}
\end{equation} 
Note that ${\bf \Psi}(n\pi)={\bf \Psi}^n(\pi)$.
Calculations of $u_p$ and $\alpha$ require knowledge of a single cell of the periodic part of the potential. Details of how to compute $u_p$ and $\alpha$ are found in reference \cite{T19b}.

A connection between two fundamental solutions of the Schr\"{o}dinger equation is formulated by a matrix equation involving a constant matrix. For example, the two fundamental 'exterior' solutions ${\bf \Psi}_{L,R}(x)$ are related by
\begin{equation}
{\bf \Psi}_{L}(x) = {\bf \Psi}_{R}(x) {\bf \Omega}, \label{genLR}
\end{equation}
where ${\bf \Omega}$ is an $x$-independent matrix.  ${\bf \Omega}$ can be determined at any matching point, say $x=n\pi$. 
This gives the relation
\begin{equation}
{\bf \Psi}_{L}(x) = {\bf \Psi}_{R}(x) \left[{\bf \Psi}_R^{-1}(n\pi){\bf \Psi}_{L}(n\pi)\right]. \label{matchn}
\end{equation}
 The matrix value ${\bf \Psi}_{L}(n\pi)$
can be expressed in terms of ${\bf \Psi}(x)$ in (\ref{PsiM}) for the periodic part of the potential. A matching between ${\bf \Psi}_{L}(x)$ and  ${\bf \Psi}(x)$ at $x=0$ yields
\begin{equation}
{\bf \Psi}_{L}(x)={\bf \Psi}(x)\left[{\bf \Psi}^{-1}(0){\bf \Psi}_{L}(0)\right].\label{match0}
\end{equation}

By combining  (\ref{matchn}) and (\ref{match0}) with the use of (\ref{LRPsi0}) and (\ref{PsiMn}), one finds
the matrix $\bf\Omega$ as
\begin{equation}
{\bf \Omega} = \left(\begin{array}{cc} k^{-1/2} & k^{-1/2}\\    \rmi k^{1/2} & - \rmi k^{1/2} \end{array}\right)^{-1} \left(\begin{array}{cc}
   \cos n\alpha & u_p^2 \sin n\alpha \\ 
   -u_p^{-2} \sin n\alpha & \cos n\alpha
  \end{array}\right)\left(\begin{array}{cc} k^{-1/2} & k^{-1/2}\\    \rmi k^{1/2} & - \rmi k^{1/2} \end{array}\right), \label{OMf}
\end{equation}
i.e.
\begin{equation}
{\bf \Omega} = \left(\begin{array}{cc} \Delta^* &\Lambda \\   \Lambda^* & \Delta \end{array}\right),\;\; \det {\bf \Omega} = 1,,\;\;|\Delta|^2=1+|\Lambda|^2, \label{OMLam}
\end{equation}
with
\begin{equation}
\Lambda =\frac{\rmi}{2}\left(\left(ku_p^2\right)^{-1}-ku_p^2\right)\sin n\alpha,\;\;
\Delta = \cos n\alpha -\frac{\rmi}{2}\left(\left(ku_p^2\right)^{-1}+ku_p^2\right)\sin n\alpha.
\end{equation}
The quantity $\Lambda$ is expressed with the use of $J(E)=\left(\left(ku_p^2\right)^{-1}-ku_p^2\right)/2$ in (\ref{LambdaLP}).
With known exact elements in ${\bf \Omega}$, scattering boundary conditions (\ref{leftB}) and (\ref{rightB}) can be re-interpreted in terms of amplitude-phase quantities. In the left asymptotic region of $x$ the amplitude-phase  solution $\Psi^{(-)}_{L}(x)$ behaves as \cite{T05a}
\begin{equation}
\Psi^{(-)}_{L}(x) \sim  \frac{1}{\sqrt{k}} {\rm e}^{-\rmi kx},
\;\;\mbox{as} \;\;x \rightarrow -\infty.
\label{psiL}
\end{equation}
$\Psi^{(-)}_{L}(x)$ corresponds, {\it via} equation (\ref{genLR}), to an equivalent expression in terms of $\Psi^{(\pm)}_{R}(x)$, given by
\begin{equation}
\Psi^{(-)}_{L}(x)=\Lambda\Psi^{(+)}_{R}(x) + \Delta \Psi^{(-)}_{R}(x),
\label{psimatch}
\end{equation}
where 
\begin{equation}
\Psi^{\pm}_{R}(x) \sim  \frac{1}{\sqrt{k}} \; {\rm e}^{\mp \rmi kn\pi}\;{\rm e}^{\pm\rmi kx},
 \;\;x \rightarrow +\infty.
\label{psiR}
\end{equation}
From (\ref{psimatch}) and  (\ref{psiR}) follows
\begin{equation}
\Psi^{(-)}_{L}(x) \sim   \frac{\Lambda}{\sqrt{k}} {\rm e}^{-\rmi  kn\pi} {\rm e}^{\rmi kx}   
+ \,\,\,  
\frac{\Delta}{\sqrt{k}} {\rm e}^{\rmi  kn\pi}{\rm e}^{- \rmi kx}, \,\,\, x \rightarrow + \infty.
 \label{psiRR}
\end{equation}
Normalizing  (\ref{psiRR}) to agree with condition (\ref{rightB}),
the transmission and  reflection  amplitudes appear as 
\begin{equation}
t = \rme^{ -\rmi  kn\pi}\frac{1}{\Delta},\;\;r = {\rm e}^{-2\rmi kn\pi}\frac{\Lambda}{\Delta}.
\end{equation}
The  transmission and reflection 
coefficients defined in (\ref{RTdef}) can be expressed in terms of $\Lambda$ as
\begin{eqnarray}
T =  {\frac{1}{1+|\Lambda|^{2}}},\;\;R=  {\frac{|\Lambda|^{2}}{1+|\Lambda|^{2}}}.
\label{Tdeff1}
\end{eqnarray}
This brief derivation is generalized to include exterior potentials with more general amplitude-phase methods in \cite{T20a} (to be published elsewhere).

\section{Concluding remarks} 
A formally exact amplitude-phase approach is explored in the context of one-dimensional scattering and Floquet/Bloch bands. The relevance of Floquet/Bloch theory is illustrated for a multi-well potential with $n=6$. Band/gap structures for a given potential explains structures of transmission bands for multi-well potentials even for $n=6$. Band types can be classified by an intrinsic quantity $J(E)$, which may have a zero or not in the band. Energy peaks of total transmission are caused by two intrinsic quantities: $J(E)$, a slowly varying functions of energy; and the intrinsic phase $\alpha(E)$, a rapidly vaying function of energy.


\section*{References}
\begin{thebibliography}{99}



\bibitem{T05a} K.-E. Thylwe, J. Phys. A: Math. Gen. {\bf38}   (2005) 235.

\bibitem{T19b} K.-E. Thylwe,  Phys. Scr. 94 (2019) 065201; https://doi.org/10.1088/1402-4896/ab40d3.

\bibitem{Griffiths01} D. J. Griffiths and C. A. Steinke, American Journal of Physics 69  (2001) 137; https://doi.org/10.1119/1.1308266.

\bibitem{Dharani16} M. Dharani, and C. S. Shastry,
AIP Conference Proceedings 1731, 110017 (2016); https://doi-org./10.1063/1.4948038

\bibitem{Shao} Z. Shao and W. Porod, Phys. Rev. B 51 (1995) 1931.

\bibitem{Oh99} G.-Y. Oh, arxiv.org/abs/cond-mat/9902181
	
\bibitem{Maiz15} F. Maiz, Physica B: Condensed Matter, 463 (2015) 93.

\bibitem{Nanda06} J. Nanda, P. K. Mahapatra, C. L. Roy,
Physica B: Physics of Condensed Matter, Vol.383(2) (2006) 232

\bibitem{Mukhopadhyay12} S. Mukhopadhyay, R. Biswas, C. Sinha,
Physics Letters A,  Vol.376(15) (2012) 1306.

\bibitem{Yu90} K. W. Yu, Computers in Physics 4, 176 (1990), https://doi.org/10.1063/1.168361.

\bibitem{Sprung93} D. W. L. Sprung, Hua Wu, and J. Martorell, American Journal of Physics 61 (1993) 1118, https://doi.org/10.1119/1.17306.

\bibitem{Bar05} D. Bar, International Journal of Theoretical Physics, Vol. 44, (2005) 1281, DOI: 10.1007/s10773-005-4686-x

\bibitem{Duke69}  C. B. Duke, {\it Tunneling in Solids} (Academic, New York and London, 1969). 

\bibitem{Mandrˆ14} S. Mandr\`{a}, J. Schrier, M. Ceotto, 
The journal of physical chemistry. A,  Vol.118(33) (2014) 6457.




\bibitem{ReviewsG} M. Dragomana, D. Dragoman,  Progress in Quantum Electronics 33 (2009) 165Ð214;
A. H. Castro Neto, F. Guinea, N. M. R. Peres, K. S. Novoselov  and A. K. Geim,  Rev. Mod. Phys. 81 (2009) 109Ð62.


\bibitem{Zubarev} A. Zubarev and D. Dragoman, Physica E 44 (2012) 1687.;A. Zubarev and D. Dragoman, J. Phys. D: Appl. Phys. 47 (2014) 425302.

\bibitem{DiazG}
D. S. D'az-Guerrero, L. M. Gaggero-Sager, I. Rodr'guez-Vargas and O. Sotolongo-Costa,
Panchadhyayee, Pradipta
Philosophical Magazine, (2013) 1. 

\bibitem{Cury88} L. A. Cury, N. Studart, 
Superlattices and Microstructures, Vol.4(2)  (1988)  245.

\bibitem{Karavaev93} G. Karavaev, N. Chuprikov,
Russian Physics Journal,  Vol.36(8) (1993) 749.

\bibitem{Kumar18} S. Kumar; S. Kumari
Int. J. of Nanoparticles, Vol 10 (2018) 92.

\bibitem{Pyykko88} P. Pyykk\"{o}, Chem. Rev. 88 (1988) 563.

\bibitem{Pereyra98} P. Pereyra,  J. Phys. A 31 (1998) 4521 .

\bibitem{Siddhant15} Siddhant Das,
American Journal of Physics 83 (2015) 590; https://doi.org/10.1119/1.4916834.

\bibitem{Zhao16}  R. Zhao, Y. Zhang, Y. Xiao, and W. Liu, J. Chem. Phys. 144 (2016) 044105. 

\bibitem{Roy} C L Roy,
Journal of Physics: Condensed Matter, Vol.5(41) (1993) 7701.

\bibitem{Chen16} C.H. Chen, P. Tseng, W.J. Hsueh, Physics Letters A 380 (2016) 2957.

\bibitem{T20a} K.-E. Thylwe,  http://arxiv.org/pdf/2005.11695.

 \bibitem{Hill} W. Magnus and S. Winkler 1979 {\it Hill's Equation} (Dover, New York).
 
 \bibitem{Bloch} F. Bloch, Z. Phys. 52 (1929) 555-600.
 
\bibitem{Grimshaw} R. Grimshaw 1990 {\it Nonlinear Ordinary Differential Equations.}
Applied Mathematics and Engineering Science Texts, Oxford: Blackwell.10:  0632027088 

\bibitem{Kittel} C. Kittel, 1996 {\it Introduction to Solid-State Physics}, 7th edition (John-Wiley, Singapore) pp. 173-196.

\bibitem{kev:pert}  J. Kevorkian and J. Cole 1981 {\it Perturbation
Methods in applied Mathematics.} Berlin: Springer-Verlag.

\bibitem{hay:oscil} {C. Hayashi} 1964 {\it Nonlinear Oscillations in Physical
Systems.} New York: McGraw-Hill.

\bibitem{Pinney}  E. Pinney,  Proc. Am. Math. Soc. 1 (1950) 681.
\bibitem{Milne} W. E. Milne,    {Phys. Rev.} {\bf 35} (1930) 863. 
\bibitem{Wheeler} J. A. Wheeler, Phys. Rev. 52 (1937) 1123.
\bibitem{Wilson} Wilson H A 1930, {\em Phys. Rev.} {\bf 35}, 948.
\bibitem{Young} Young H A 1931, {\em Phys. Rev.} {\bf 38}, 1612.;
Young H A 1932, {\em Phys. Rev.} {\bf 39}, 455.
\bibitem{APbound} K.-E. Thylwe, J. Math. Cem. {\bf 53}, (2015) 1608.; K.-E Thylwe, Phys. Scr. {\bf85} (2012) 065009.
\bibitem{T18b} K.-E. Thylwe, J Math Chem 56 (2018) 2674.

\end{thebibliography}
\end{document}